\documentclass[reprint, amsmath, amssymb, aps, floatfix, prl, superscriptaddress]{revtex4-2}

\usepackage{graphicx}% Include figure files
\usepackage{dcolumn}% Align table columns on decimal point
\usepackage{bm}% bold math
\usepackage{gensymb} % for \degree symbol
\usepackage{hyperref}% add hypertext capabilities
\usepackage{orcidlink} % for \orcidlink command
\begin{document}

%\preprint{APS/123-QED}

\title{\textbf{Signals too small to sense: Physical and information-theoretic limits to induction-based magnetoreception in birds} 
}% 

\author{Daniel R. Kattnig\orcidlink{0000-0003-4236-2627}}
 \email{d.r.kattnig@exeter.ac.uk}
 \affiliation{
 Department of Physics and Astronomy, University of Exeter, Stocker Road, Exeter,EX4 4QL, United Kingdom
}
\affiliation{
Living Systems Institute, University of Exeter,\\ Stocker Road, Exeter, EX4 4QD, United Kingdom
}

% \author{Daniel R.\ Kattnig\,\orcidlink{0000-0003-4236-2627},$^{1,2,*}$ \\\\
% \normalsize $^{1}$ Department of Physics and Astronomy, University of Exeter, Stocker Road, Exeter, EX4 4QL, United Kingdom. \\\\
% \normalsize $^{2}$ Living Systems Institute, University of Exeter, Stocker Road, Exeter, EX4 4QD, United Kingdom. \\\\
% \normalsize $^{*}$ Electronic mail: d.r.kattnig@exeter.ac.uk\\\\
% }

\date{\today}% It is always \today, today,

\begin{abstract}
A recent study [\emph{Science} 2025, eaea6425] proposes that magnetoreception in pigeons may arise from electromagnetic induction within the semicircular canals of the inner ear. In this framework, motion through the geomagnetic field is suggested to generate an induced electromotive force that leads to ion redistribution in the endolymph, activation of voltage-gated calcium channels, and subsequent engagement of downstream neural circuits. In this work, we examine the physical plausibility of this mechanism using a toy model of the induction process combined with an information-theoretic analysis. We find that, under idealised assumptions, Faraday induction in the semicircular canals would not generate a signal of sufficient informational content to support the extraction of directional magnetic field information from the geomagnetic field. However, the model supports the possibility of inferences due to radio-frequency (RF) electromagnetic waves of a miniscule amplitude, thereby providing a potential rationalisation of their disruptive effect on avian compass navigation. We stress that our analysis does not call into question the experimental evidence for magnetically responsive pathways within the vestibular system of pigeons. Rather, it constrains the class of viable physical mechanisms, indicating that a functionally competent magnetosensory system likely relies on different sensing principles or, if induction-based, on a different sensing architecture, while highlighting induction as a potential interference pathway of RF electromagnetic fields. 
\end{abstract}


\maketitle

%\tableofcontents

\section*{\label{sec:into} Introduction}

Many vertebrates, in particular migratory birds, show robust behavioural evidence for magnetoreception--the ability to detect and respond to the Earth’s magnetic field \cite{Johnsen2005}. Yet, despite decades of research, the biophysical basis of this sense remains unresolved. Several principal hypotheses have been proposed, including a magnetite-based mechanism \cite{Winklhofer2010,Shaw2015}, a radical-pair mechanism (including a direct light-dependent mechanism \cite{Hore2016} and a dark or indirectly light-dependent mechanism \cite{Denton2024}), and an induction-based mechanism that uses electroreception to infer a time-dependent magnetic field \cite{Jungerman1980,Paulin1995}.

Recently, a renewed induction-based hypothesis has gained attention following experimental and anatomical studies in pigeons (\emph{Columba livia}). In two papers \cite{Nimpf2019,Nordmann2025}, Keays and colleagues proposed that the electrolyte-filled semicircular canals of the inner ear lagena organs, which are central to vestibular sensing, might also serve as magnetic sensors, by harnessing electromagnetic induction. The idea has been supported by the identification of a population of specialised type II hair cells in the semicircular cristae expressing a sensitive voltage-gated ion channel and whole-brain activity mapping revealing light-independent, magnetically induced neuronal activation in brain regions connected to the vestibular system.

The core idea draws on classical electromagnetic induction, as described by Faraday's law \cite{Feynman2006Vol2}, as the sensing principle: a changing magnetic flux through a loop, such as produced by the loop rotating in a magnetic field, produces an electromotive force (EMF), equating to a solenoidal electric field in the circumference of the loop (viewed from a loop-fixed observed). Nimpf \emph{et al.}~proposed that the required induction loops are provided by the three roughly orthogonal semicircular ducts in the inner ear \cite{Nimpf2019,Winklhofer2019}. These conduits are filled with an electrolyte solution, endolymph (essentially a $150\mskip3mu$mM aqueous KCl solution \cite{Money1966}), and divided by the cupula (a gelatinous glycoprotein hydrogel) in the ampullae, which supposedly (cf.\ \cite{Watanabe2001}) forms an electrically insulating barrier. The EMF, dropping predominantly over the cupula, is thought to give rise to a charge separation at either side of the cupula, producing a potential difference across the cupula and the underlying sensory epithelium at its base. The hypothesis posits that this potential difference is large enough to be sensed by electroreceptive hair cells, apparently expressing a splice isoform of a voltage-gated calcium channel (CaV1.3) also involved in electroreception in certain fish, i.e., elasmobranchs \cite{Kalmijn1971,Kalmijn1982,Winklhofer2019}. If the bird thus rotates its head relative to the geomagnetic field, the change in flux through the canal loop could enable a dynamic magnetoreception, whereby, according to the proposers of the hypothesis, vestibular inputs could be distinguished from magnetic inputs due to orthogonal directions that maximise their perception (cf.\ Fig.\ 5 in \cite{Nordmann2025}), provided that additional visual cues are integrated as a reference.

We will argue here that, while it is possible that magnetic stimuli might be perceived via the inner ear, as demonstrated in \cite{Wu2011,Nimpf2019,Nordmann2025}, a goal-orientated extraction of magnetic field information, i.e., functionally competent magnetoreception, from electric induction in the semicircular loops is physically impossible. If magnetic information is truly sensed via the inner ear in pigeons, which appears to be well supported by the body of work, it must involve a different sensing circuit if induction-based or a different sensing paradigm. This conclusion has been reached on the basis of a toy model of the sensory circuit, a ring-shaped structure comprising an electrolyte and a dielectric diaphragm, which, in essence, forms a RC-circuit. We calculate the induced potential across the ring circuit and demonstrate that the sensory receptive voltage difference is by at least three orders of magnitude smaller than the inherent noise of the sensing circuit, rendering active perception impossible, regardless of subsequent signal processing or sensory pooling. An information-theoretical bound is used to demonstrate that the rate of information extractable from such a noisy system is insufficient to underpin the notion of functionally competent magnetoreception, regardless of downstream neural summation or processing. We also show that the spatial distribution of the induced charge separation is incompatible with the sensing through hair cells because the induced changes are necessarily equalised on a length scale, the Debye screening length, small in comparison to a single voltage-gated ion channel. Furthermore, incomplete blockage of the ion conduction by the cupula, i.e., acting as a leaky seal, is shown to drastically reduce the induced potential difference. While the model is simple, the relevant parameters can be well estimated, or have been chosen to maximise the effect, demonstrating that no trivial modification can alter the conclusion reached here. However, we will argue that the suggested induction-based mechanism could provide a rationalisation for the disruptive effects of radio-frequency (RF) electromagnetic waves in avian magnetoreception, which are commonly attributed to the radical-pair mechanism, but are not quantitatively explainable by this model \cite{Ritz2004}.

\section*{\label{sec:model} A model for electro-magnetoreception in ring structures}

\subsection*{\label{sec:model:defs}Definitions and governing equations}

\begin{figure}[tb]
\centering
\includegraphics[width=0.4\textwidth]{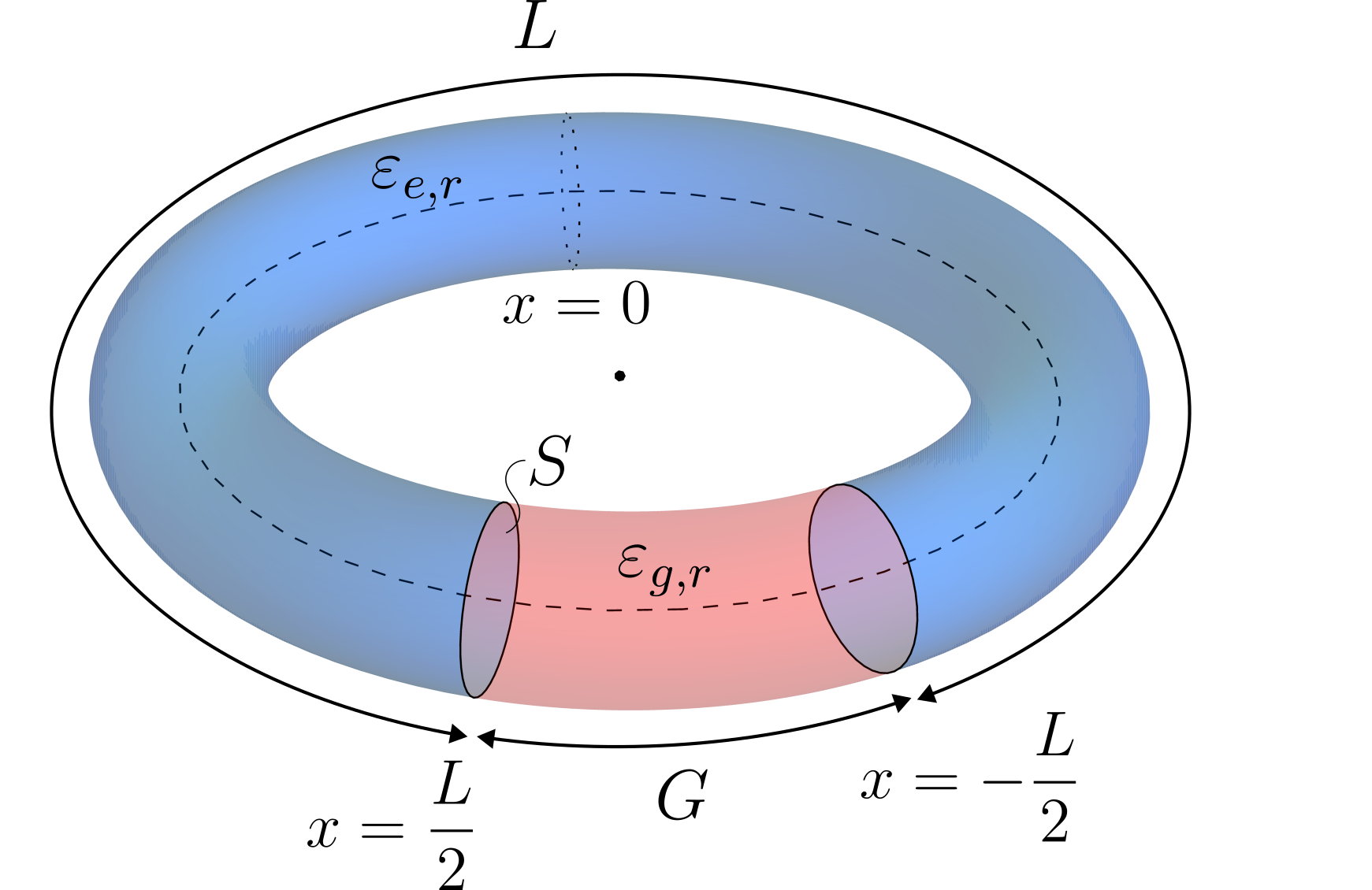}
\caption{Summary of the toy model of induction-based magnetoreception via a dielectric-gapped electrolyte ring. \label{fig:ring_model}}
\end{figure}

We model the toroidal ``semicircular'' channel of the vestibular apparatus as a ring of diameter $d$ and circumference $U=\pi d$ (see Fig.\ \ref{fig:ring_model}). This yields a description in terms of a single coordinate, the polar angle $\theta$, which we map onto $x = \tfrac{d}{2} \theta$, which runs along the arc of the ring from $x=-\tfrac{1}{2} U$ to $x= \tfrac{1}{2} U$. The ring is taken to comprise a gap of length $G$, modelled after the gelatinous cupula, and an electrolyte of length $L$ (such that $U = L + G$), representative of the endolymph in the semicircular channels. Here, we assume that the gap is dielectric, imposing a physical barrier to circumferential ion flow, which maximises the induced potential difference; in the Supporting Information, we extend the discussion to the scenario of non-zero ion conductivity. For convenience, we take the electrolyte to be centred on $x = 0$, i.e., extending from $x=-\tfrac{1}{2} L$ to $x= \tfrac{1}{2}L$. The permittivities in the gap and the electrolyte are denoted $\varepsilon_g = \varepsilon_{g,r}\varepsilon_0$ and $\varepsilon_\ell = \varepsilon_{\ell,r}\varepsilon_0$, respectively.

A time-varying magnetic field $\mathbf{B}(t)$ passing through the ring creates an electromotive force $\mathcal{E}$ and associated electric field $E_{\mathrm{ind}}$, which in the rest frame of the ring is given by
\begin{equation}
    \mathcal{E} = \oint \mathbf{E}_{\mathrm{ind}} \cdot d\mathbf{l} = -\frac{d\Phi_B}{dt}, \label{eq:emf}
\end{equation}
where $\Phi_B$ is the magnetic flux over the ring (Faraday's law of induction) \cite{Feynman2006Vol2}. Assuming that the cross-sectional radius of the canal duct is much smaller than the ring diameter $d$ (which holds true here; \emph{vide infra}), for a thin ring in a $B$-field uniform over the cross-section, $\mathbf{E}_{\mathrm{ind}}$ is tangential to the ring with
\begin{equation}
    E_{\mathrm{ind}}(t) = -\frac{d}{4} \frac{dB}{dt}. \label{eq:Eind}
\end{equation}
$E_{\mathrm{ind}}(t)$ acts as an external driving force on ions in the electrolyte, inducing their redistribution and thereby giving rise to an opposing polarisation field $E_{\mathrm{pol}}$. Unlike the solenoidal induced field, the resulting $E_{\mathrm{pol}}$ is a conservative field, expressible as a gradient of the electrostatic potential $\varphi(x,t)$,
\begin{equation}
    E_{\mathrm{pol}}(x,t) = -\frac{\partial}{\partial x} \varphi(x,t)
\end{equation}
and obeying
\begin{equation}
\oint \mathbf{E}_{\mathrm{pol}} \cdot d\mathbf{l} = 0.
\end{equation}
The total electric field inside the ring is thus the sum of the induced field and the electrostatic field created by ion redistribution
\begin{equation}
    E(x,t) = E_{\mathrm{ind}}(t) + E_{\mathrm{pol}}(x,t).
\end{equation}
The ion dynamics in the electrolyte is governed by the Poisson-Nernst-Planck (PNP) equations \cite{Bazant2004}: The flux density $J_i$ for species $i$ including diffusion and electromigration is given by
\begin{equation}
    J_i = -D_i \frac{\partial c_i}{\partial x} + \frac{z_i D_i F}{RT} c_i E(x,t), \label{eq:pnp_flux_Ji}
\end{equation}
where $c_i$ is the molar concentration, $D_i$ the diffusion coefficient, $z_i$ the charge number, and all other quantities have their canonical meaning. Note that we neglect advection because magnetoreception is supposed to involve motions around axes within the ring plane that do not induce vestibular output through fluid motion \cite{Nordmann2025}. Using the continuity equation $\frac{\partial c_i}{\partial t} = -\frac{\partial J_i}{\partial x}$ and eq.\ \eqref{eq:pnp_flux_Ji}, the equation of motion of the particle densities is obtained as
\begin{equation}
    \frac{\partial c_i}{\partial t} = \frac{\partial}{\partial x} \left[ D_i \frac{\partial c_i}{\partial x} - \frac{z_i D_i F}{RT} c_i \left(E_{\mathrm{ind}}(t) - \frac{\partial \varphi}{\partial x} \right) \right],
\end{equation}
whereby the electric potential within the electrolyte is related to the ion distribution by Poisson's equation
\begin{equation}
    -\frac{\partial^2 \varphi}{\partial x^2} = \frac{\rho(x)}{\varepsilon_\ell},
\end{equation}
with the charge density given by $\rho(x) = F \sum_i z_i c_i$. As the barriers at $x=-\frac{L}{2}$ and $x=\frac{L}{2}$ are here assumed to be impermeable to ions, the total flux of each species must be zero at these interfaces:
\begin{equation}
    J_i\left(\pm \tfrac{L}{2}, t\right) = 0.
    \label{eq:boundary_J}
\end{equation}
The charge redistributions induced by the reorienting geomagnetic field are small, allowing for a perturbation theoretical expansion that considers $E_{\mathrm{ind}}(t)$ as a perturbation, thereby facilitating the derivation of simple expressions for the charge density and the potential across the ring. These details are provided in the Appendix.

\subsection{\label{sec:sec:model:defs} Expressions central to induction-based magnetoreception}

We here state results of the theoretical analysis that are central to the subsequent discussion; detailed derivations are provided in the Appendix and Supporting Information. For this purpose, we specialise the discussion to a 1:1 electrolyte of singly-charged ions, and use subscripts + and - to denote the cation and anion, respectively, i.e., $z_{+} = - z_{-} = 1$.  We will also assume $D_{+} = D_{-} = D$ for simplicity. As endolymph features potassium chloride at a concentration far exceeding other ions \cite{Money1966} and the diffusion coefficients of potassium and chloride ions agree closely (transport numbers for K$^{+}$ and Cl$^{-}$ are roughly 0.49 and 0.51)     \cite{Vanysek1992,Robinson2002}, these assumptions do not limit the suitability of the model to assess the fundamental feasibility of induction-based magnetoreception.

We assume that the magnetic field varies sinusoidally, i.e., employing phasors, $B(t) = \tilde{B}_0 e^{i\omega t}$. The induced electric field resulting from Faraday's law, eq.\ \eqref{eq:Eind}, is then given by $E_{\mathrm{ind}}(t) = \tilde{E}_0 e^{i\omega t}$, with $\tilde{E}_0 = -i\omega \tfrac{d}{4} \tilde{B}_0$. Subject to the assumptions from above, the linearised dynamics of this system is amenable to an analytical treatment, with details summarised in the Appendix. Here, we state the expressions obtained for the linearised charge density, $\rho(x,t) = F (c_+(x,t) - c_-(x,t)) \approx F (\delta c_+(x,t) - \delta c_-(x,t))$, with $\delta c_i = c_i - c_0 + O(E_{\mathrm{ind}}^2)$ denoting the first-order deviations from the bulk concentration $c_0$, and the corresponding induced potential drops across the gap and the electrolyte, which are central to the Discussion.

The spatial behaviour of this system is characterised by $\gamma$, which for harmonic driving is given by
\begin{equation}
    \gamma^2 = \kappa^2 + \frac{i\omega}{D}. \label{eq:gamma_osc}
\end{equation}
The charge density induced in the electrolyte is obtained as $\rho(x,t) = \tilde{\rho}(x) e^{i\omega t}$ with
\begin{eqnarray}
    \tilde{\rho}(x) = A \sinh\left(\gamma x \right), \label{eq:rho_driven}
\end{eqnarray}
where $A$ is an integration constant depending on geometric and dynamics parameters; it is stated explicitly in eq.\ \eqref{eq:A} in the Appendix. The voltage drop across the electrolyte, which is equal to the maximal potential difference measurable across the electrolyte, is obtained as $V_{\ell}(t) = \tilde{V}_{\ell} e^{i\omega t}$ with
\begin{equation}
    \tilde{V}_{\ell} = \tilde{E}_{\mathrm{ind}}L - A \left[ \frac{i\omega L \cosh(\Lambda)}{\varepsilon_{\ell} D \kappa^2 \gamma} + \frac{2 \sinh(\Lambda)}{\varepsilon_{\ell} \gamma^2} \right], \label{eq:Ve_driven}
\end{equation}
with $\Lambda = \gamma\tfrac{L}{2}$.

If the magnetic field drive has instead an exponential time dependence, $B(t) = \tilde{B}_0 e^{-t/\tau}$, giving rise to the induced electric field $E_{\mathrm{ind}}(t) = \tilde{E}_0 e^{-t/\tau}$, where $\tilde{E}_0 = -\frac{d}{4\tau} \tilde{B}_0$, an approach analogous to the above can be used, but employing an exponential ansatz for the time dependence of dynamic quantities, e.g.\ $\rho(x,t) = \tilde{\rho}(x) e^{-t/\tau}$ and $V_{\ell}(t) = \tilde{V}_{\ell} e^{-t/\tau}$. The solutions are then found to be of the same form as above, except that the characteristic wavenumber $\gamma$ is given by
\begin{equation}
    \gamma^2 = \kappa^2 - \frac{1}{D\tau}.
    \label{eq:gamma_exp}
\end{equation}
Thus, eqs.\ \eqref{eq:rho_driven} and \eqref{eq:Ve_driven} apply in this scenario with redefined $\gamma$.

\section{Results and Discussion}

Using the model established above, we evaluated the central quantities of the hypothesis. The model parameters were chosen as follows. We used a semicircular duct diameter of $5\mskip3mu$mm, corresponding to the typical size reported in \cite{Nimpf2019}. The cross-section of the endolymph-filled lumen was estimated using the average membranous duct radius of pigeons reported in \cite{Hullar2006} ($170\mskip3mu\mu$m).

The electrolyte parameters were chosen to model endolymph, which contains potassium chloride as the main electrolyte ($154\mskip3mu$mM; the concentration of other ions is small, e.g.\ Na$^+$: $0.91\mskip3mu$mM) and which at body temperature ($T = 316\mskip3mu$K) has a dynamic viscosity comparable to that of water ($\sim1\mskip3mu$mPa~s) \cite{Money1966}. Therefore, we assumed an aqueous solution of $154\mskip3mu$mM KCl. The cationic and anionic diffusion coefficients were assumed equal and set to $D = 2 \times 10^{-9}\,\mathrm{m^2/s}$, consistent with the measured diffusion coefficients of aqueous solutions at room temperature \cite{Vanysek1992}. The room-temperature diffusion coefficient provides a conservative estimate, as a higher $D$ would allow charge perturbations to relax even faster. The relative permittivity of the electrolyte was assumed to be equal to that of water ($\varepsilon_{\ell,r} = 78.4$).

The cupula was initially modelled as a perfect dielectric, completely blocking the ion conductance (cf.\ eq.\ \eqref{eq:boundary_J}), in agreement with the suggestion of \cite{Nimpf2019,Nordmann2025}. This assumption maximises the induced potential difference and charge separation, but contradicts some evidence in pigeons supporting the view that the endolymphatic fluid passes through the subcupular space or the top of the crista ampullaris even for small applied pressure gradients, as they would result from head rotations \cite{Watanabe2001,Damiano1999}. The literature on other species also strongly supports the fact that the cupula in both the semicircular canals and the mechanosensory lateral lines (the evolutionary precursor to the inner ear in fish and amphibians) is highly permeable to ions and has electrical conductivity nearly identical to that of the surrounding fluid. Using energy dispersive X-ray spectrometry, Flock mapped the distribution of potassium and chloride ions in freeze-dried inner ear tissue, finding high ion contents in the cupula of the semicircular canal and concluding that the tectorial membrane and the cupula are completely permeable to the surrounding electrolyte \cite{Flock1977}. Jielof et al. demonstrated that the gelatinous lateral line in fish and amphibians has an electrical conductivity essentially identical to that of the surrounding electrolyte \cite{Jielof1952}, which means that it acts as a conductor rather than a dielectric barrier. Rabbitt et al.\ have discussed the poroelastic nature of the cupula, modelling it as a porous mesh where the endolymph (and its constituent ions) can flow through the subcupular space and the gel matrix itself when subjected to pressure gradients or electrical fields \cite{Rabbitt2004}. For these reasons, we have later included simulations that involve a residual ion conductance in the gap region. 

The dielectric properties of the cupula have not been studied. However, being described as a poroelastic gelatinous structure largely composed of water, the dielectric properties of gelatin-water gels appear to provide a reasonable reference \cite{Fricke1940}. The static dielectric constant of such gels can reach several times that of water; it peaks for 45\% gelatin, for which $\varepsilon_{r}\approx 500$ \cite{Fricke1940}. Given the uncertainty associated with this parameter, we decided to study a wide range, including the apolar limit ($\varepsilon_{g,r} = 4$), intermediate values ($\varepsilon_{g,r} = 100$), and the highly polar limit ($\varepsilon_{g,r} = 1000$), as would arise from the Maxwell-Wagner-Sillars effect or interfacial polarisation from mobile counter-ions getting trapped at the microscopic gel-water interfaces. The width of the cupular was systematically varied or fixed to have a width $G$ equal to the channel diameter ($340\mskip3mu\mu$m).

\subsection*{Induced charge separation and potential}

\begin{figure}[tb]
\centering
\includegraphics[width=0.35\textwidth]{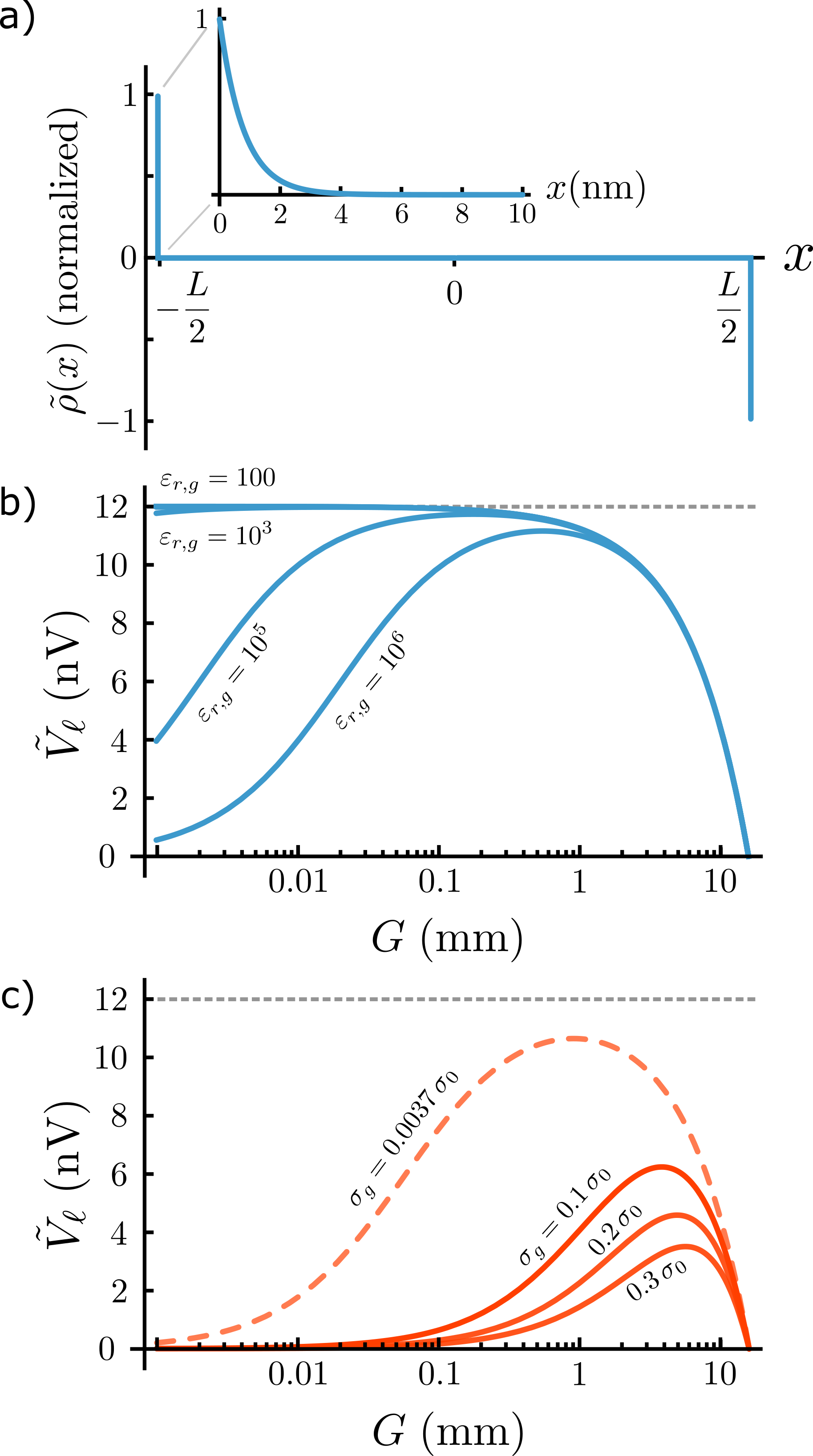}
\caption{Charge distribution $\tilde{\rho}$, a), and induced potential difference over the electrolyte $\tilde{V}_{\ell}$, b) and c), of a harmonically driven dielectric-electrolyte ring. a) gives the normalized charge accumulation over the entire electrolyte and, as an insert, within the nanometre-sized boundary zone at the electrolyte-dielectric interface. b) and c) illustrate $\tilde{V}_{\ell}$ as a function of the width of the cupula width $G$, in b) for the idealised non-conducting cupula of variable relative dielectric constant $\varepsilon_{r,g}$ as indicated; and in c) for the conductive cupula of $\varepsilon_{r,g}=100$ with conductivity $\sigma_g$ as indicated. The red dashed line in c) uses the conductivity of isoelectric gelatin gel. The cupula is actually expected to have electrical conductivity essentially identical to that of the surrounding electrolyte, i.e.\ $\sigma_g\approx\sigma_0$. The details of the calculation and parameters used are specified in the text. The grey dashed lines indicate the total electromotive force. \label{fig:results}}
\end{figure}

First, we assumed a time-dependent magnetic field with harmonic characteristics, as could result from head scans. The angular velocity was set to $700^\circ$/s, in agreement with the maximal velocity reported for pigeons in free flight \cite{Kano2018}; and the magnetic field strength was taken as $50\,\mu$T, appropriate for the geomagnetic field at moderate latitude in Europe. Despite highly speculative and contradicting the current knowledge on cupulas, we focused on the ideal case of a perfect dielectric barrier, as this maximises the effect. Equation~\eqref{eq:rho_driven} was used to calculate the charge density along the ring for the selected $\varepsilon_{g}$. The maximal ion concentration difference between cations and anions, $\rho/F$, at the interface was found to range from $1.6 \times 10^{-14}$ mol/L to $4.1 \times 10^{-12}$ mol/L as $\varepsilon_{g,r}$ increased from 4 to $10^3$. The increase with $\varepsilon_{g}$ is in line with eq.~\eqref{eq:A} in the limit of low frequencies and mesoscopic $L$ (i.e., $\omega \rightarrow 0$ and $\tanh(\Lambda) \rightarrow 1$) and common intuition, that the interface charge increases with gap capacitance and thus $\varepsilon_{r,g}/G$. The peak amplitude lagged the $B$-field by approximately $90^\circ$, as expected for a magnetic field varying slowly relative to intrinsic ion dynamics (cf.\ Eq.~\eqref{eq:gamma_osc}). These changes are evidently very small, reaching at most $2.6 \times 10^{-11}$ of the bulk concentration $c_0$ in the high $\varepsilon_{g,r}$-limit, and are also highly localised. The accumulation of ions is maximal at the interface and decays rapidly, with the characteristic decay length given by the Debye length $\gamma^{-1} \approx \lambda_D = 0.80$ nm. Thus, ion enrichment/depletion is restricted to a region significantly narrower than a single calcium channel (diameter $\sim 10\mskip3mu$nm \cite{Schwenzer2025}). Beyond a few $\lambda_D$, the electrolyte is homogeneous; only at the opposite interface does a symmetric and equally small counterion accumulation manifest (see Fig. \ref{fig:results}a). The small size of the charge separation becomes more evident when calculating the absolute amount of ions ``accumulated": for $\varepsilon_{r,g} = 100$, it formally amounts to 0.018 ions (which physically manifests as a minute shift in the probability of an ion occupying the boundary layer). 

The charge separation is accompanied by a potential difference across the electrolyte, which opposes the induced EMF. As the voltage profile attains its extremal values at the electrolyte–gap interfaces, we calculate $\Delta V_\ell = \varphi \left(+\tfrac{L}{2}\right) - \varphi \left(-\tfrac{L}{2}\right)$ as a measure of signal strength. This represents the maximum potential difference generated within the electrolyte as a result of induction. For the perfectly dielectric gap with $G = 340\mskip3mu\mu$m, $|\Delta V_\ell|$ is found to be $11.7\mskip3mu$nV, practically independent of $\varepsilon_{r,g}$ in the range of 4 - 1000. In this scenario, the electric field within the electrolyte is essentially nullified as a consequence of its high conductivity and low driving frequency, with the induced polarisation field nearly entirely compensating the solenoidal field. Therefore, the induced potential difference is close to the  total EMF of $12.0\mskip3mu$nV, the difference reflecting the distribution of the EMF over the ring and the gap. Hence, this scenario is close to optimal with respect to sensing magnetic fields via induced potential differences. 

Figure~\ref{fig:results}c shows the dependence of $|\tilde{V}_\ell|$ on the gap width $G$ for a range of $\varepsilon_{g,r}$. It is evident that deviations from the ideal scenario described above are only observed in the highly polar limit ($\varepsilon_{g,r} \gtrsim 10^5$). Although such high $\varepsilon_{g,r}$ are commonly found in biological soft tissues with a high water content, they are dominated by ionic diffusion processes near cell membranes and are not expected to be relevant here \cite{Gabriel1996}. Even if this was the case, the predicted gap width would yield a $|\Delta V_\ell|$ close to the maximum, thereby still allowing most of the EMF to be observed in the electrolyte. 

Above, we have assumed that the gelatinous cupula is electrically insulating to maintain the maximal voltage drop across the sensory epithelium. However, experiments on pigeons suggest that the endolymphatic fluid can pass through the subcupular space or the top of the crista ampullaris, ultimately leading to leakage and short-circuiting of the induced potential. It is also expected that gelatinous biological matrices generally exhibit non-negligible ionic conductivity and that surrounding tissues form parallel conductive pathways, which provide additional routes for charge dissipation, reducing or eliminating the voltage gradient required for transduction. To assess these effects, we have extended the model to include conductivity in the gap region; details are summarised in the appendix. Figure~\ref{fig:results}b illustrates the effect of non-zero gap conductivity for $\varepsilon_{r,g} = 100$. A gap conductivity of only 10\% of the electrolyte conductivity (as would, e.g., result from a radial gap in the seal of 9$\mskip3mu\mu$m width) is predicted to reduce $|\Delta V_{\ell}|$ to $1.1\mskip3mu$nV for $G=340\mskip3mu\mu$m; even larger conductivities would  practically eliminate the induced potential, whereby an increase in $G$ could counteract the reduction to some extent. The intrinsic conductivity of gelatin-gels, on the other hand, would only reduce the potential difference to $8.8\mskip3mu$nV provided that the seal is otherwise perfect, at least if the conductivity of $80\cdot10^{-6}\mskip3mu\Omega^{-1}$cm$^{-1}$ applying to isoelectric gelatin gel with 30 \% gelatin in water is used \cite{Fricke1939}. However, it should be noted that added salts could strongly increase the conductivity and a 
gelatinous hydrogel swollen in a $150\mskip3mu$mM KCl solution has a conductivity almost identical to the bulk $150\mskip3mu$mM KCl solution \cite{Amsden1998,Bard2001}. In summary, while induced potential differences on the order of the EMF are theoretically achievable for idealised parameters, realising these is critically dependent on an intact seal. Under realistic leaky conditions, the electrolyte potential differences remain significantly smaller than the EMF, likely by at least an order of magnitude. For the not unlikely conductivity of $\sigma_g\approx\sigma_0$ the induced potential difference vanishes.

\subsection{The noise floor}

The existence of inducible potential differences on the order of $10\mskip3mu$nV has been taken to support the plausibility of induction-based magnetoreception in the inner ear \cite{Nordmann2025}. Admittedly, some fish have been reported to be able to detect electric fields as small as $5\mskip3mu$nV\,cm$^{-1}$ \cite{Kalmijn1971,Kalmijn1982} using long ampullary canals (ampullae of Lorenzini; $\sim 10\mskip3mu$cm and highly resistive canal walls specifically evolved to prevent current leakage \cite{Clusin1977}), which is clearly remarkable (for example, the basic restrictions for human exposure at the relevant frequency limit the internal field strength to $0.26\mskip3mu$Vm$^{-1}$ \cite{ICNIRP2010}, equating to an induced EMF of $3.5\mskip3mu$mV across the ring sensor). However, the described ring sensor is unlikely to be able to detect Earth's magnetic fields varying at rates consistent with animal motion. We follow the general idea of Jungman and Rosenblum to argue this point \cite{Jungerman1980}, but adapt the discussion to the specific sensor model at hand and then apply an information theoretical bound to suggest impossibility.

The sensor can be described as an $RC$-circuit, combining an Ohmic resistor, i.e., the electrolyte, with a plate capacitor formed by the interfaces and the dielectric gap. This circuit is inherently noisy: thermal agitation of charge carriers within the conductor at equilibrium causes voltage and current noise, known as Johnson–Nyquist noise \cite{Johnson1928,Nyquist1928}. This thermal noise is fundamental, i.e., present in all electrical circuits regardless of applied voltage, and will drown out weak signals, thereby limiting the sensitivity of any electrical measuring instrument. The mean-square noise voltage of an $RC$ circuit is
\begin{equation}
    \overline{V_{n,d}^{2}} = 4k_{\mathrm{B}}T\, R_d\,\Delta \nu_d = \frac{k_{\mathrm{B}}T}{C_d},    
\end{equation}
where $R_d$ is the resistance, $C_d$ the capacitance and $\Delta \nu_d$ the bandwidth of the circuit. The second equality follows from the first by substituting the equivalent noise bandwidth of a $RC$-circuit, $\Delta \nu_d = 1/(4R_dC_d)$. The subscript $d$ denotes the detector.

The relevant electrical properties of the detector can be estimated by assuming a toroidal shape with cross-section $S$. The resistance $R_d$ arises from the finite conductivity of the electrolyte column of length $L$ and is given by
\begin{equation}
    R_d = \frac{L}{\sigma_0 S}. \label{eq:resistance}
\end{equation}
The capacitance of the system results from the dielectric slit and the double-layer capacitances associated with the accumulation of ions at the electrolyte–barrier interfaces. Assuming parallel-plate capacitors and using the Debye length $\lambda_D = \kappa^{-1}$ as the characteristic separation for the double-layer capacitances, the series combination of individual capacitances yields
\begin{equation}
    C_{\mathrm{total}} = \left( \frac{G}{\varepsilon_g S} + \frac{2}{\varepsilon_{\ell} \kappa S} \right)^{-1}.
\end{equation}
Together, these expressions provide an estimate of the root-mean-square (RMS) noise voltage, $V_{\mathrm{rms}} = \sqrt{\overline{V_{n,d}^{2}}}$, and the signal-to-noise ratio of the detector,
\begin{equation}
    \left(\frac{S}{N}\right)_d = \frac{|\tilde{V}_\ell|^2}{\overline{V_{n,d}^{2}}},
\end{equation}
expressed in terms of energy densities. From Eq.~\eqref{eq:resistance} we obtain $R_d = 78\mskip3mu$k$\Omega$, while the capacitance varies strongly with polarity, ranging from $9.5\times 10^{-15}$ to $2.4\times 10^{-12}$ F for $\varepsilon_{g,r} = 4$--$1000$. The RMS noise voltages range from $670\,\mu$V to $43\,\mu$V, reflecting the widely different detector bandwidths that arise for different $\varepsilon_{g,r}$, which range from $340\mskip3mu$MHz to $1.4\mskip3mu$MHz. Clearly, the detector noise greatly exceeds the voltage drop induced across the electrolyte.

However, this does not necessarily invalidate the mechanism, as the signal is expected to reside entirely in a narrow low-frequency band. Assuming that the signal is transmitted for cognitive processing through a channel of lower bandwidth, $\Delta \nu_s$,
\begin{equation}
    \overline{V_{n,s}^{2}} = \frac{\Delta \nu_s}{\Delta \nu_d}\,\overline{V_{n,d}^{2}}
    = 4k_{\mathrm{B}}T\,R_d\,\Delta \nu_s = N_0\,\Delta \nu_s,
\end{equation}
where the subscript $s$ refers to the complete detection system (detector plus its low-bandwidth, signal-selective output channel), and $N_0$ is the noise spectral density. The resulting RMS noise amplitude is independent of the dielectric properties and equal to $36.8\sqrt{\Delta \nu_s}\,\mathrm{nV\,s^{1/2}}$.

Thus, for a plausible bandwidth of $1\mskip3mu$kHz--suitable for detecting induction changes during rapid head rotation--the signal-to-noise power ratio is predicted to be $1.0\times 10^{-4}$ for the idealised induced $|V_{\ell}|$ ($\varepsilon_{g,r} = 4$--$1000$), and correspondingly smaller for the lower $|V_{\ell}|$ associated with more highly polar gaps. For the leaky gap with a gap conductivity of $0.1 \sigma_0$, the SNR is $8.2 \times 10^{-7}$; for the sensor that assumes a perfect seal but residual gelatin conductivity, the  SNR is $5.8 \times 10^{-5}$. These estimates suggest that the signal will be overwhelmed by noise even for optimal, yet unrealistic, sensor parameters and strongly bandwidth-limited detection \cite{Weaver1990,Adair1991,Adair2000}.

Could the signal be recovered from the noise by further reducing the system bandwidth? This is impossible because magnetoreception based on a time-dependent induction signal inherently requires detection on a commensurate timescale. To make this argument quantitative, we consider the Shannon–Hartley theorem, which limits the rate at which information can be transmitted over a communications channel of bandwidth $\Delta\nu$ in the presence of additive white Gaussian noise \cite{MacKay2003,Bialek1987}:
\begin{equation}
    K = \Delta\nu \, \log_{2}\!\left(1+\frac{S}{N}\right),
\end{equation}
where $K$ is the channel capacity (bits per second). This gives the maximum amount of error-free information per unit time that can be transmitted at a given bandwidth in the presence of noise. In the limit of low $S/N$ with $N = \Delta \nu N_0$, applicable here,
\begin{equation}
    K \approx \frac{1}{\ln 2} \left(\frac{S}{N_0}\right),
\end{equation}
i.e., the channel capacity becomes independent of bandwidth.

Induction-based magnetoreception—inferring magnetic field orientation—requires detecting the orientation for which the moving sensor produces a characteristic induction signal (e.g.\ a maximum, minimum, or fastest change), i.e., the signal must be linked to the field direction. Assume that the sensor rotates with angular velocity $\omega$ and discriminates $n_{\phi}$ distinct orientations over a scan of 180$^{\circ}$ (e.g., \ $n_{\phi} = 36$ for a resolution of 5$^{\circ}$ that birds can resolve). For each orientation, the induction signal must be measured to identify the characteristic direction. If each measurement is represented by $b$ bits, the required channel rate for magnetoreception becomes
\[
K_{\mathrm{req}} = b \, n_{\phi} \, \frac{\pi}{\omega}.
\]
For a 5$^{\circ}$ resolution and even a very coarse 4-bit signal quantisation (16 levels), one obtains $K_{\mathrm{req}} > 560\,\mathrm{s^{-1}}$. However, the practically achievable $K$ falls decisively short of this value. In fact, for the perfectly sealed dielectric gap with $\varepsilon_{r,g}=4$--$1000$, we find $K = 0.15$ s$^{-1}$, while the weakly leaky model with $\sigma_G = 0.1 \sigma_0$ provides $K = 0.022$ s$^{-1}$ only (despite taking into account the reduced impedance of the conducting device, which gives rise to a reduced noise density).

It is evident that the actual channel capacity is far too small to support functional magnetoreception, even if an unrealistic model with negligible cupula conductivity is employed. In the low-$S/N$ limit, this conclusion is independent of the system bandwidth: although reducing the bandwidth lowers the noise, it also proportionally restricts the information extraction rate, preventing any improvement when the information rate itself is the decisive quantity, as it is in a hypothetical rotation-based induction sensor. We also note that pigeons have been reported to be able to recognise magnetic anomalies associated with variations of magnetic field on the order of $100\mskip3mu$nT, suggesting that not only the geomagentic field, but fractions thereof, are quantifiable, necessitating a larger bit-rate than assumed here (e.g., 9 bit to distinguish $100\mskip3mu$nT on top of $50\mskip3mu\mu$T, suggesting $K_{\mathrm{req}} > 1300$) \cite{Wiltschko2009}. 

Based on these observations, we conclude that functional induction-based magnetoreception utilising the semicircular canals in pigeons, as suggested, is untenable. No post-processing can overcome this limitation, and no sensor element (e.g.\ ion channel), regardless of its sensitivity, can alter the conclusion that the required information is simply not available at a sufficient rate. This does not preclude the existence of other induction-based mechanisms, including in pigeons or other animals, but their sensing paradigm and/or physical parameters must differ substantially from the model proposed in \cite{Nimpf2019,Nordmann2025} to overcome the fundamental limits outlined here. However, it is not inconceivable that the suggested induction processes could be perceived by the animal in a non-functional (or disruptive; \emph{vide infra}) way, in particular if magnetic field intensities of larger amplitude and/or fast transients or high frequencies are involved. The experiments in \cite{Nimpf2019,Nordmann2025} use a $150\mskip3mu\mu$T-magnetic field jump rotating by $6^{\circ}$. Assuming that the switching time is limited by the time-constant of the Helmholtz coil system employed, this configuration is expected to exponentially switch $15.7\mskip3mu\mu$T in $12\mskip3mu$ms, yielding a somewhat larger $|V_\ell|$ of $25\mskip3mu$nV and maintaining $K\approx0.7\mskip3mu$s$^{-1}$ employing eq.\ \eqref{eq:gamma_exp} for the idealised, non-leaky scenario, which is still roughly 1000 times too slow to meet the $K_{\mathrm{req}} > 560\mskip3mu$s$^{-1}$ threshold for navigation.

\subsection{Beyond the rotating geomagnetic field}

\begin{figure}[tb]
\centering
\includegraphics[width=0.4\textwidth]{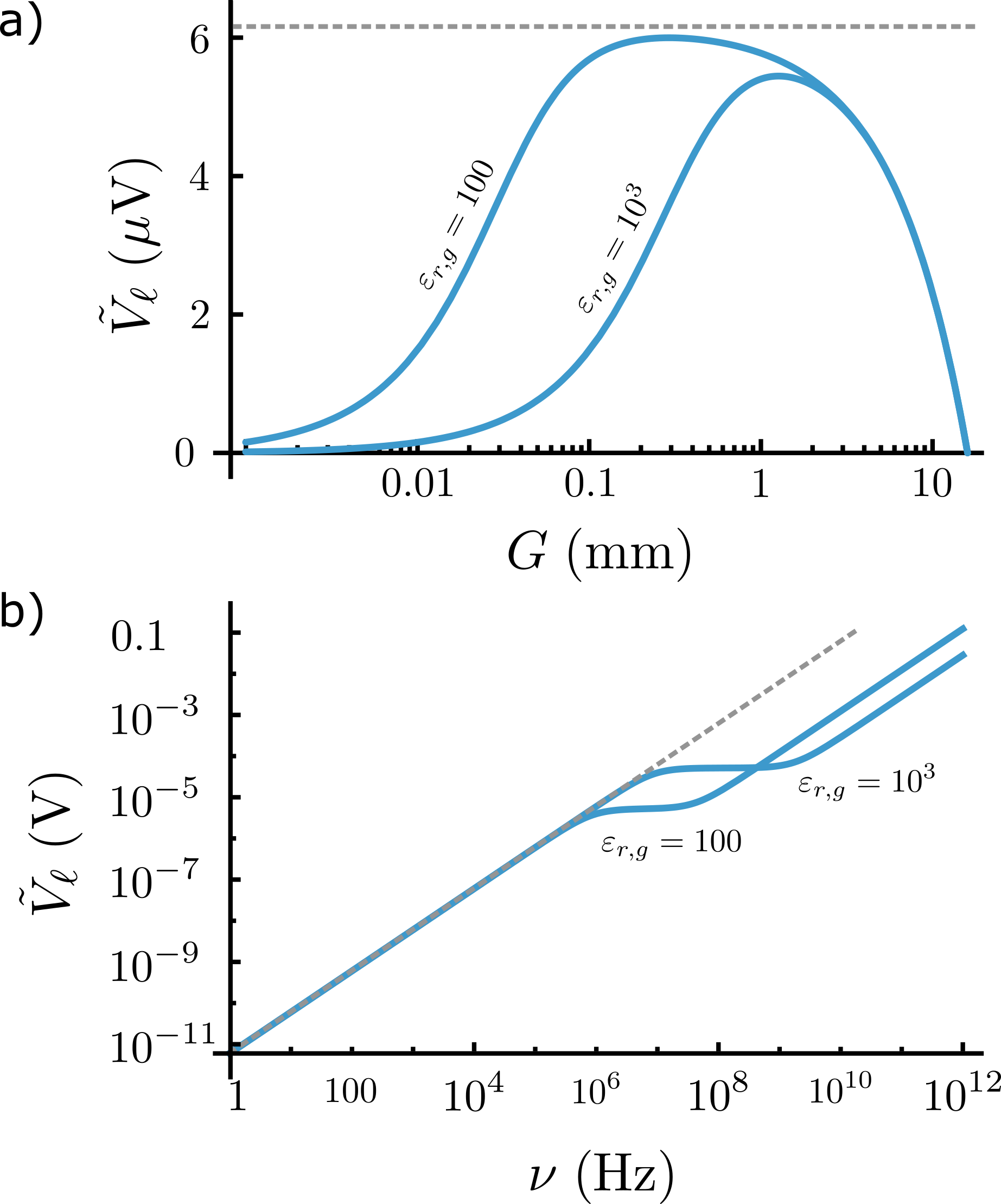}
\caption{Induced potential difference across the electrolyte $\tilde{V}_{\ell}$ for the ring sensor driven by rf-electromagnetic fields. a) $\tilde{V}_{\ell}$ for a $1\mskip3mu$MHz electromagnetic field of $50\mskip3mu$nT amplitude as a function of the width of the cupula $G$, idealized to be non-conducting. b) $\tilde{V}_{\ell}$ for $G = 340\mskip3mu\mu$m as a function of the rf frequency, with other parameters as for a). Results are shown for two relative dielectric constants of the cupula, as indicated on the graphs.  The grey dashed lines indicate the total electromotive force. \label{fig:results_rf}}
\end{figure}

The induced EMF scales with the time derivative of the magnetic flux through the sensing loop (Faraday's law of induction). As a direct consequence, induction-based magnetoreception is, in principle, inherently sensitive to time-varying, i.e. electromagnetic, fields. As illustrated in Fig.\ \ref{fig:results_rf}, the induced potential difference across the electrolyte increases with the frequency of an oscillatory magnetic field, reflecting the proportionality to $\dot{\Phi}_B \propto \omega$. This implies that even extremely weak oscillatory fields can generate induced potentials that exceed those predicted for geomagnetic-field detection via head rotation or comparable quasi-static stimuli. The effect becomes particularly pronounced in the MHz regime and above, where the rapid temporal variation strongly enhances the induced voltage. For example, a field of frequency $10\mskip3mu$MHz and an amplitude of only $1\mskip3mu$nT induces $|\tilde{V}_{\ell}| = 1.6\mskip3mu\mu$V in the non-lossy sensor with $\varepsilon_{g,r}=100$, far exceeding the geomagnetic-induced potential.

If such high-frequency signals are to be transduced by ion channels, they must first overcome a fundamental biophysical hurdle: the cell membrane acts as a severe low-pass filter. Modelled as a parallel RC circuit, the capacitance of the lipid bilayer effectively short-circuits transmembrane voltages at frequencies above the kilohertz range; MHz-frequency-induced potentials largely bypass the membrane \cite{Kotnik2000}. However, transduction could remain physically plausible to high frequencies through non-linear rectification \cite{Astumian1995,Wust2020}. Because asymmetric ion channels act similar to electrical diodes, the inherent asymmetry of the open pore's current-voltage characteristics could ``demodulate'' an oscillatory RF signal into a net DC shift, even though macroscopic channel gating kinetics cannot track MHz frequencies. Indeed, selected experimental studies of the interactions of RF electromagnetic fields with isolated ion channels suggest that this rectification in the $1\mskip3mu$MHz -- $1\mskip3mu$GHz  range is feasible in principle \cite{Ramachandran2010,Ramachandran2011,Kim2011,Dharia2011,Wust2020}. This phenomenon has been suggested to arise from asymmetric ion conductivity combined with characteristic ion dwell times of the order of $10\mskip3mu$ns, leading to measurable changes in net ion currents under RF exposure. Ultimately, this demodulation mechanism could allow high-frequency AC fields to incrementally alter the transmembrane potential. Barsoum and Pickard found that the cell membranes of the algae Chara braunii and Nitella flexilis showed rectification properties at frequencies up to $10\mskip3mu$MHz \cite{Barsoum1982}, and the International Commission for Non-Ionising Radiation Protection (ICNIRP) guidelines assert that induced electric fields with frequencies up to $10\mskip3mu$MHz can stimulate nerves (in humans and for internal electric fields exceeding those explored here) \cite{Saunders2007,ICNIRP2020}. In view of the observations and the exquisite sensitivity of the pigeon's voltage-gated calcium channel, the extension of a putative induction-based magnetosensation to higher frequencies is at least not inconceivable. 

These considerations are especially relevant in the broader context of avian magnetoreception \cite{Hore2016}. RF magnetic-field effects on the avian compass are frequently discussed as diagnostic signatures of the radical pair mechanism (RPM), although the observations remain difficult to reconcile quantitatively within that framework \cite{Ritz2004,Hiscock2017,Leberecht2023}. An induction-based contribution offers a complementary perspective: RF magnetic fields might modulate lagena-related sensing pathways, thereby perturbing not the spin dynamics central to the RPM itself, but rather the lagena-derived reference signal used for inclination-based measurements. In this scenario, RF interference would indirectly degrade compass performance by destabilising the reference frame against which radical-pair signals are interpreted. The observed frequency window (approximately $100\mskip3mu$kHz to $120\mskip3mu$MHz in birds \cite{Leberecht2023}) could result from the need to induce a significant EMF at the lower bound and the ability to rectify high-frequency stimuli through ion channels at the upper limit.

Finally, these arguments highlight an important experimental implication. Investigations of induction-based magnetoreception must carefully control for RF magnetic-field contamination, including high-frequency ripple from switch-mode power supplies and other electronically generated noise sources. Even weak RF components, if within the relevant frequency range, could produce disproportionately large induced potentials and thereby confound measurements of low-frequency or quasi-static magnetic-field responses.

\section*{Conclusions}

The analysis presented here demonstrates that the induction-based model of magnetoreception in the avian inner ear, as suggested in \cite{Nimpf2019,Nordmann2025}, when examined using quantitative electromagnetics and information-theoretic limits, cannot provide a physically viable signal. The constraints imposed by induced potentials, thermal noise, and the fundamental limits of information transmission preclude the semicircular canals from functioning as reliable geomagnetic field sensors under realistic conditions. While one cannot categorically rule out that evolutionary adaptations might somehow circumvent these barriers, doing so would require a sensing architecture fundamentally different from the macroscopic induction loop proposed.

We note that this analysis is not meant to diminish the compelling biological observations that indicate that a magnetically responsive pathway exists within the vestibular system of the pigeon \cite{Wu2011,Nimpf2019,Nordmann2025}. Rather, our results show that its physical basis cannot reside in the classical Faraday induction of low-frequency rotating magnetic fields within the semicircular canals. On the other hand, the hypothesised induction mechanism is recognised as potentially sensitive to RF electromagnetic fields and, while unfit as compass, could provide an explanation for their disruptive effect on avian magnetoreception (a phenomenon currently widely attributed to the radical pair mechanism).

This situation is not unusual in the history of sensory biology or, in fact, science: mechanisms that appear plausible at first glance have often, under closer physical scrutiny, yielded to deeper and more unexpected explanations \cite{Meister2016}. That the presently proposed mechanism fails on first principles should therefore not be regarded as a refutation of the biological evidence, but as an invitation to search for an alternative, perhaps more subtle, origin of the effect, capable of reconciling both empirical observations and physical constraints. In this sense, the gap identified here is an opportunity. By clarifying what magnetoreception cannot be, we hope to have helped narrow the space of what it might be \cite{Johnsen2005}. Progress will come from a sustained dialogue between physics and biology—neither discipline alone is sufficient, but together they can reveal mechanisms that are both physically sound and biologically verified.

\begin{acknowledgments}

The author thanks Prof.\ Michael Winklhofer (University of Oldenburg) and Prof.\ Peter Hore (University of Oxford) for their comments on an early draft of this manuscript. This research did not receive dedicated funding. 

\end{acknowledgments}

\appendix

\section{Appendixes}

\subsection*{Linearisation of the Poisson--Nernst--Planck equations}

To obtain the first-order response to a weak induced electric field, we write the local ionic concentrations as small deviations from their uniform equilibrium value \(c_0\),
\[
c_i(x,t) = c_0 + \delta c_i(x,t).
\]
Substituting this into the PNP equation and retaining only linear terms in \(\delta c_i\) and \(\phi\), we obtain
\[
\frac{\partial \delta c_i}{\partial t}
= D_i \frac{\partial^2 \delta c_i}{\partial x^2}
+ \frac{z_i D_i F c_0}{RT}
\frac{\partial^2 \phi}{\partial x^2}.
\]
Here we have used that $E_{\mathrm{ind}}$ is spatially uniform, i.e., $\frac{\partial E_{\mathrm{ind}}}{\partial x} = 0$, and discarded products such as \(\delta c_i\,\partial_x \phi\) or \(E_{\mathrm{ind}} \,\partial_x \delta c_i\), which are second order, i.e., the polarisation potential and \(\delta c_i\) are of first order as induced by $E_{\mathrm{ind}}(t)$. The linearised no-flux boundary condition at the dielectric interfaces becomes
\[
\left. \frac{\partial \delta c_i}{\partial x} \right|_{x=\pm\tfrac{L}{2}}
= \frac{z_i F c_0}{RT}
\left. \left( E_{\mathrm{ind}}(t) - \frac{\partial \phi}{\partial x} \right) \right|_{x=\pm\tfrac{L}{2}}.
\label{eq:boundary_deltac}
\]
Finally, Poisson’s equation relates the perturbations of concentration to the electric potential, allowing the potential term to be decouple from the above expressions
\[
\frac{\partial^2 \phi}{\partial x^2}
= - \frac{F}{\varepsilon_{\ell}}
\sum_i z_i \, \delta c_i.
\]

We now specialise to a symmetric \(1{:}1\) electrolyte with \(z_+=-z_-=1\). Define charge- and salt-density perturbations \(\rho = F(\delta c_+ - \delta c_-)\) and \(\sigma = \delta c_+ + \delta c_-\), where subscripts $+$ and $-$ denote the cation and anion, respectively. Assuming \(D_+=D_- = D\), \(\sigma\) evolves independently,
\[
\frac{\partial \sigma}{\partial t}
= D \frac{\partial^2 \sigma}{\partial x^2}.
\]
For the zero-flux boundary conditions and a system started at uniform equilibrium, i.e., $\sigma(x,0) = 0$, this equation is trivially solved by $\sigma(x,t) = 0$, which implies the anti-symmetry of the concentration deviations $\delta c_+(x,t) = -\delta c_-(x,t)$.

Subtracting the continuity equations and using Poisson’s equation gives
\[
\frac{\partial \rho}{\partial t}
= D\left(
\frac{\partial^2 \rho}{\partial x^2}
- \kappa^2 \rho
\right),
\]
where we have introduced $\kappa^2 = \frac{2 F^2 c_0}{\varepsilon_{\ell} R T}$, which is related to the inverse Debye length $\lambda_D = \kappa^{-1}$ and the ionic conductivity $\sigma_0 = \varepsilon_{\ell} \kappa^2 D$. Perturbations of the charge density thus relax with characteristic time constant $\tau_D = (D\kappa^2)^{-1}$. The corresponding boundary condition becomes
\begin{equation}
\left. F(J_+ - J_-) \right|_{x=\pm\tfrac{L}{2}}
= \left. \sigma_0 E - D \frac{\partial \rho}{\partial x} \right|_{x=\pm\tfrac{L}{2}}
= 0.
\label{eq:boundary_rho}    
\end{equation}

\subsection*{Harmonic driving fields}

Let the magnetic field be sinusoidal, expressed using phasors as \(B(t) = \tilde{B}_0 e^{i\omega t}\), which produces \(E_{\mathrm{ind}}(t) = \tilde{E}_0 e^{i\omega t}\) with \(\tilde{E}_0 = - i\omega \frac{d}{4}\tilde{B}_0\). We seek solutions of the form \(\rho(x,t) = \tilde{\rho}(x)e^{i\omega t}\), with analogous expressions assumed for the electric field $E(x,t)$ and the potential $\phi(x,t)$, the amplitude functions likewise denoted by the tilde. Inserting this ansatz into the linearised charge-density equation yields
\[
\frac{d^2 \tilde{\rho}}{dx^2}
- \left(\kappa^2 + \frac{i\omega}{D}\right)\tilde{\rho}
= 0.
\]
Introducing
\begin{equation}
    \gamma^2 = \kappa^2 + \frac{i\omega}{D},
\end{equation}
the general solution of this differential equation can be represented as $\tilde{\rho}(x) = A_1 \sinh(\gamma x) + A_2 \cosh(\gamma x)$. Given the symmetry of the problem and noting that the forcing field is uniform over the ring, the charge distribution will be inversion symmetric, with positive and negative charges accumulating at opposite dielectric barriers. Therefore, we choose the solution of the form
\begin{equation}
    \tilde{\rho}(x) = A \sinh\left( \gamma x \right),
\end{equation}
with $A = A_1$ denoting an integration constant to be determined from boundary conditions. Specifically, boundary condition eq.\ \eqref{eq:boundary_rho} requires
\begin{equation}
   \sigma_0  \tilde{E}\left(\frac{L}{2}\right) = D \left. \frac{\partial \tilde{\rho}}{\partial x} \right|_{x=\tfrac{L}{2}} = D A \gamma \cosh(\Lambda), \label{eq:bc2}
\end{equation}
where $\Lambda = \gamma\tfrac{L}{2}$. The constant electric field in the dielectric, $E_g$, then follows using the continuity of the electric displacement field normal to the interface: $\varepsilon_{\ell} \tilde{E}\left(\frac{L}{2}\right) = \varepsilon_g \tilde{E}_g$. On the other hand, from Gauss' law, $\frac{d E_{\mathrm{pol}}}{dx} = \frac{\rho}{\varepsilon_{\ell}}$, $\tilde{E}_{\mathrm{pol}}(x)$ can be obtained by integrating $\tilde{\rho}$. Combining $E_{\mathrm{pol}}$ and $E_{\mathrm{ind}}$, we thus obtain
\begin{equation}
    \tilde{E}(x) = \tilde{E}_{\mathrm{pol}}(x) + E_{\mathrm{ind}} = \frac{A}{\varepsilon_{\ell} \gamma} \cosh(\gamma x) + C_1,
\end{equation}
where $C_1$ is another integration constant. These equations and the additional requirement that the total field integrates to the EMF, eq.\ \eqref{eq:emf}, allow fixing all constants.
Specifically, the charge density amplitude $A$ is finally obtained as
\begin{equation}
  A = \frac{\mathcal{E} \, \varepsilon_g \, \kappa^2}{G \, \gamma \cosh\left(\Lambda\right)} \left[ 1 + \frac{\varepsilon_g}{\varepsilon_{\ell} G} \left( L \frac{i\omega}{D \gamma^2} + \frac{2 \kappa^2}{\gamma^3} \tanh\left(\Lambda\right) \right) \right]^{-1} \label{eq:A} 
\end{equation}
and
\begin{equation}
    C_1 = A \left[ \frac{i\omega \cosh(\Lambda)}{\varepsilon_{\ell} D \kappa^2 \gamma} \right].
\end{equation}
The potential subject to the reference $\tilde{\phi}(0) = 0$ is then given as
\begin{equation}
    \tilde{\phi}(x) = -\frac{A}{\gamma^2 \varepsilon_{\ell}} \sinh (\gamma  x) - (C_1 - E_{\mathrm{ind}})\, x,
\end{equation}
and the potential difference across the electrolyte as provided in eq.\ \eqref{eq:Ve_driven}  follows.

\subsection*{The leaky diaphragm}

For a non-zero conductivity of the cupula, we solve the linearised Poisson-Nernst-Planck equations in both regions, using an ansatz of the form
\begin{equation}
     \tilde{\rho}(x) = A \sinh\left( \gamma_\ell x \right)\mspace{20mu}\text{for}|x|\leq\tfrac{L}{2}
\end{equation}
for the electrolyte (as above), and
\begin{equation}
     \tilde{\rho}(x') = A' \sinh\left( \gamma_g x' \right)\mspace{20mu}\text{for}|x'|\leq\tfrac{G}{2}
\end{equation}
for the ion conductive cupula, where $x'=x-\tfrac{L+G}{2}$ is measured from its centre. The constants $A$ and $A'$ and additionally ensuing integration constants can be found by solving a linear system constructed to fulfil the following boundary conditions: the conservation of the total current density, the continuity of the electric displacement at the interface, the continuity of the ion concentration at the interface, and the Faraday's law condition, eq.\ \eqref{eq:emf}. The resulting expressions are impractically complex and have been evaluated numerically.

% \nocite{*}

\bibliographystyle{apsrev4-2} 
\bibliography{references}% Produces the bibliography via BibTeX.

\end{document}